# Angles in the SI: treating the radian as an independent, unhidden unit does not require the redefinition of the term 'frequency' or the unit hertz

Paul Quincey

National Physical Laboratory[1], Hampton Road, Teddington, TW11 0LW, United Kingdom.
paul.quincey@npl.co.uk

**Abstract**

Some recent papers have argued that frequency should have the dimensions of angle/time, with the consequences that 1 Hz = 2π rad/s instead of 1 s$^{-1}$, and also that $v = \omega$ and $h = \hbar$. This letter puts the case that this argument redefines the quantity 'frequency' and then draws conclusions from equations that rely on the standard definition being used. The problems that this redefinition is designed to address arise from the widespread unstated adoption of the Radian Convention, which treats the radian as a dimensionless quantity equal to the number 1, in effect making the radian a 'hidden' unit. This convention is currently 'built-in' to the SI, when it should be separable from it. The unhelpful status of angles in the SI can be remedied with minimal disruption by (1) changing the definition of the radian from 'a radian equals 1 m/m' to e.g. 'a right angle equals π/2 radians', and (2) acknowledging that the adoption of the Radian Convention is acceptable, when made explicit. The standard definitions of 'frequency' and the hertz should remain unchanged.

1. Introduction

Are the correct SI units for frequency s$^{-1}$ or rad/s? When answering this, a great deal of time could be saved by realising that this is not a philosophical question about the physical nature of frequency, but instead a simple question of how we choose to define the term 'frequency'. The definition of terms may seem dull to most working scientists, but it is a necessary and important part of science, especially when setting out the basic equations of physics or assigning units to physical quantities. There is a formal, international process for defining scientific terms, specifically involving the ISO 80000 series of standards. These are not infallible or immutable, but changes to them should not be proposed without very good reasons. Because of the wide acceptance of these definitions, we refer to them below as the 'standard' definitions.

In ISO 80000-3:2019, the definition of frequency (with symbol *f* or *v*) (item 3-17.1) is 'inverse of period duration (item 3-14)', with unit Hz or s$^{-1}$. The definition of period (*T*) (item 3-14) is 'duration of one cycle of a periodic event', unit s, where 'a periodic event is an event that occurs regularly with a fixed time interval'. These are entirely consistent, clear, and unambiguous.

---

[1] The paper expresses the views of the author and these do not necessarily reflect those of NPL Management Ltd.

And yet we now see published papers urging scientists to use rad/s as the unit for frequency [1-4]. It is entirely reasonable, for wave phenomena, to define a quantity *related* to the frequency that is the 'rate of change of the phase angle', with unit rad/s, but to call this 'frequency', and to give it the symbol $v$, simply invites confusion. This quantity is in fact defined in ISO 80000-3:2019 (item 3-18), and is called 'angular frequency' with the symbol $\omega$.

The proposal in references [1-4] to tacitly redefine the term frequency has direct and undesirable consequences, leading to statements that '$v = \omega$' and, within quantum physics, that '$h = \hbar$', for example. This letter will firstly make the simple point that, however familiar they may be, equations relating physical quantities will no longer be correct when the quantities within them are redefined. Secondly, it will suggest the root causes of the confusion in this area, to help provide an understanding of the problems that the proposal is designed to address. Finally, it will point out a clear way forward.

## 2. The relationship between equations and how physical quantities are defined

In not-so-distant history, the concepts of mass and weight were commonly confused. Indeed this was the subject of a resolution at the 3$^{rd}$ CGPM – the highest decision-making body for metrology - in 1901. The resolution was entitled *Declaration on the unit of mass and on the definition of weight*; it considered 'the necessity to put an end to the ambiguity which in current practice still exists on the meaning of the word weight, used sometimes for mass, sometimes for mechanical force'.

The problem was not just semantic. For example, it needs to be unambiguous that the density $\rho$ of a substance is its mass $M$ divided by its volume $V$, not its weight $W$ divided by its volume. The equation is therefore $\rho = M/V$. If we measure weight, the equation should be changed to $\rho g = W/V$, where $g$ is the acceleration due to gravity. We cannot use different definitions of quantities and expect equations to remain the same.

To put this another way, if we want to know the *mass* of a given volume of water, we use the equation $M = \rho_w V$, where $\rho_w$, the density of water, has the units for mass/volume. If we want to know the *weight* of a given volume of water, we should use the equation $W = g\rho_w V$. It would be very confusing, though not incorrect, to use the equation $W = \rho_w V$, where $\rho_w$ is still called the 'density' of water, but represents its weight per unit volume. And it would be plainly wrong to say that because $W = \rho_w V$ is a valid equation, and $M = \rho_w V$ is also a valid equation, we have proved that $W = M$, and so mass and weight are the same thing after all. We need to be aware that the density symbol $\rho_w$ in the two equations is defined differently in each case.

### 3. The root causes of the confusion about frequency: '1 hertz = 1 cycle per second', and the Radian Convention

There are perhaps two reasons why problems with frequency units have arisen and remained unresolved for so long.

Firstly, there has been a reluctance to describe a hertz as an 'inverse second', rather than a supposedly more tangible 'something per second'. The ISO 80000 definition above makes it clear that the hertz is the appropriate unit for 'number of periodic events per second'. The habit for wave phenomena has been to describe the hertz as the unit for 'cycles per second', which is consistent with 'inverse second' when the term 'cycle' is a description of the relevant dimensionless 'periodic event'. However, there has been a persistent urge to interpret the term 'cycle' to mean an angle of 360° or $2\pi$ rad, which would mean that 1 Hz equals $2\pi$ rad/s, not 1 s$^{-1}$. An early example can be found in 1936 [5]. In effect, this tradition is to redefine the hertz from its standard definition, while leaving its name unchanged – a sure recipe for confusion.

The second cause is the widespread convention, currently included within the foundations of the SI, to treat plane angles as dimensionless quantities with implicit units of radians. This is described in [6] as the Radian Convention. This convention effectively 'crosses out' a constant that would appear in complete equations, those which are valid for any choice of unit. The simplification this brings to equations explains why the convention is so widespread. The constant is a plane angle with the value of 1 rad, which in [6] is called $\theta_N$. If the constant is reinserted, or at least is acknowledged to have been crossed out, dimensional relationships are clarified, and much potential confusion can be avoided. For example, the complete equation relating angular frequency to frequency is $\omega = 2\pi\theta_N v$. In this form it is clear that the unit for $\omega$ is that for $v$ multiplied by the chosen angle unit. When the $\theta_N$ is crossed out, it is easy to think, mistakenly, that $\omega$ and $v$ should have the same units, and that 'frequency' and 'angular frequency' are different versions of the same quantity.

The Radian Convention is purely presentational; it does not change the definitions of physical quantities or their natural dimensions. However, it does mean that dimensional analysis cannot be applied to equations that have adopted it. The complete equations relating photon energy $E$ to frequency and angular frequency are $E = hv$ and $E = h\omega/(2\pi\theta_N)$, where $h$ is the Planck constant. When the angle is made explicit in this way, the SI unit for $\omega$ is unambiguously rad/s, for $h$ it remains J·s, the complete unit for action, and for $h/(2\pi\theta_N)$, commonly called[2] $\hbar$, it becomes J·s/rad, the complete unit for angular momentum [6].

---

[2] As explained in reference [6], while the quantity of action $h/2\pi$ is generally called $\hbar$, the quantity of angular momentum $h/(2\pi\theta_N)$ is also called $\hbar$, because the $\theta_N$ is habitually crossed out, following the Radian Convention.

## 4. The expressions 'ν = ω' and 'ℏ = h'

As just stated, when the angle unit information that is conventionally hidden from equations is made explicit, and the standard definitions of frequency and angular frequency are used, the equation relating ν to ω is $\omega = 2\pi\theta_N \nu$. The statement that 'ν = ω' is therefore not a consequence of making the angle unit unhidden, but of redefining one or both of ν and ω from their standard definitions. The implicit proposal in [1-4] is effectively to give frequency, ν, the same definition as angular frequency, ω, so it is no surprise that these authors then conclude that 'ν = ω'. Their argument has nothing to do with restoring hidden angle units; it is purely a consequence of redefining the frequency ν.

In quantum physics, it is only by declaring 'ν = ω', and also assuming that the equations $E = h\nu$ and $E = \hbar\omega$ both remain valid, that we deduce that 'ℏ = h'. But redefining ν to equal ω will of course change the meaning of h in $E = h\nu$, much like the meaning of $\rho_w$ was changed in the example of mass and weight above. In practice, if we redefine ν to equal ω we should change this equation to $E = \hbar\nu$, and the exercise simply proves that ℏ = ℏ.

The root cause of the confusion in both these cases is the Radian Convention being adopted without realising that this has consequences. As pointed out in [6], when the Radian Convention is removed, and so the hidden angle units are made visible, $E = h\nu$ and $E = \hbar\omega$ remain valid with the standard definitions of ν and ω; h remains equal to 6.62 x 10$^{-34}$ J·s (the familiar value), and the only change is that ℏ is here equal to 1.05 x 10$^{-34}$ J·s/rad (the familiar numerical value, with the hidden angle unit for an angular momentum made visible). It is simply wrong to say that including hidden angle units leads to the conclusion that 'ℏ = h'.

## 5. Conclusion

There are genuine problems with how angles are currently treated within the SI. Apart from the situations described above, problems are apparent in quantity-handling software, for example, within which angles must be treated differently to any other quantity that has a choice of unit. The SI definitions 1 rad = 1 m/m and 1 sr = 1 m$^2$/m$^2$ also lead to algebraic nonsense such as 1 rad$^2$ = 1/(1 sr) and 1 rad = 1 s/s. These problems can all be traced to the Radian Convention being adopted without realising that this is a simplification that loses information. The convention is not wrong in itself, but the problems arise when people do not realise that the convention is adopted primarily for mathematical convenience, and it needs to be removed in circumstances when full physical information is required. The way forward in this area lies in understanding how and why the Radian Convention is applied, and when it should be removed, not in redefining other quantities in an attempt to "put things right" by other means.

Specifically, the way forward consists of two parts. Firstly, acknowledging that angles are not inherently dimensionless, as explained in [7], and replacing the current SI 'definition' of the radian, as a derived unit equal to 1 m/m, with a definition independent of other SI base units, such as 'a right angle equals π/2 rad'. If the introduction of the radian as a new SI 'base unit' were considered too radical, a new category of 'complementary unit' could be used. This reflects its status as one of the complement of five necessary base units [6], but one that is significantly different to the others because it would be the only one whose size is fixed geometrically, and will therefore never be a matter for discussion within the CIPM Consultative Committee on Units. The steradian would be a derived unit equal to 1 rad$^2$ [e.g. 8, 9].

The current definition of the radian, adopted by the SI Brochure in 1981, has meant that the Radian Convention is 'built in' to the SI, when it needs to be detached from it to avoid the problems arising. The situation is partially recognised in the current SI Brochure [10], where Section 2.3.3 states: 'However, it is a long-established practice in mathematics and across all areas of science to make use of rad = 1 and sr = 1. For historical reasons the radian and steradian are treated as derived units, as described in section 2.3.4.'

The second part consists of acknowledging that the widespread adoption of the Radian Convention in equations, when made explicit, is acceptable and will continue, while realising that it needs to be removed in software, for example, as explained more fully in [6]. This would avoid the upheaval of replacing all familiar equations involving angular quantities with unfamiliar ones containing $\theta_N$. As explained here, the standard definitions of frequency and the hertz should remain unchanged.

The two changes would create, for the first time, a unit system that is consistent with the physical world, with minimal practical disruption. Making the radian a base (or complementary) unit would be a profound step for the SI, and it would require many familiar equations, constants, and units for physical quantities to be reassessed. However, the actual changes required would be small; replacing the official SI unit for angular momentum, J·s, with J·s/rad, for example, and a realisation that some familiar 'definitive' equations like **L = r × p** are simplified versions of the complete equations, in this case **L = (r × p)**/$\theta_N$. Once the process of 'letting go' is complete, the resulting system would be inherently simpler and more logical, with angles treated exactly the same as other physical quantities. Moreover, the complex arguments about this topic that have continued for many decades could be brought to a close.